\documentclass[12pt]{article}
\usepackage{graphicx} 
\usepackage[colorlinks=true,linkcolor=blue]{hyperref} 
\usepackage{lineno}
\usepackage{xspace}
\usepackage{caption}
\usepackage{subcaption}

\newcommand\pubdate{\today}

\def\Title#1{\begin{center} {\Large #1 } \end{center}}
\def\Author#1{\begin{center}{ \sc #1} \end{center}}
\def\Address#1{\begin{center}{ \it #1} \end{center}}

\newcommand\pubblock{\rightline{\begin{tabular}{l}  \\ 
         \pubdate  \end{tabular}}}
\newenvironment{Abstract}{\begin{quotation}  }{\end{quotation}}
\newenvironment{Presented}{\begin{quotation} \begin{center} 
             PRESENTED AT\end{center}\bigskip 
      \begin{center}\begin{large}}{\end{large}\end{center} \end{quotation}}

\newcommand{\mcal}{\mathcal}
\newcommand{\mrm}{\mathrm}

\newcommand{\alphas}{\alpha_{\mrm{s}}}

\newcommand{\alphasmz}{\alpha_{\mrm{s}}\left(m_{\mrm{Z}}\right)}

\newcommand{\muR}{\mu_{\mrm{R}}}
\newcommand{\muF}{\mu_{\mrm{F}}}

\newcommand{\pt}{p_{\mathrm{T}}}

\newcommand{\et}{E_{\mathrm{T}}}
\newcommand{\vareta}{\vert \eta \vert}
\newcommand{\etgamma}{E_{\mathrm{T}}^{\gamma}}

\newcommand{\etagamma}{\eta^{\gamma}}

\newcommand{\varetagamma}{\vert \eta^{\gamma} \vert}

\newcommand{\thirdlumi}{139\,\mrm{fb}^{-1}\xspace}

\newcommand{\sherpa}{S{\scriptsize{HERPA}}\xspace}

\newcommand{\pythia}{P{\scriptsize{YTHIA}}\xspace}
\newcommand{\jetphox}{J{\scriptsize{ETPHOX}}\xspace}
\newcommand{\nnlojet}{N{\scriptsize{NLOJET}}\xspace}

\newcommand{\herwig}{H{\scriptsize{ERWIG}}\xspace}
\newcommand{\powheg}{P{\scriptsize{OWHEG}}\xspace}
\newcommand{\powhegbox}{P{\scriptsize{OWHEG}} B{\scriptsize{OX}} {\normalsize{v2}}\xspace}

\begin{document}

\begin{titlepage}

\pubblock

\vfill

\Title{Precision measurements of jet and photon \\ production at ATLAS}

\vfill

\Author{Daniel Camarero Muñoz (on behalf of the ATLAS Collaboration\footnote{\noindent Copyright 2023 CERN for the benefit of the ATLAS Collaboration. CC-BY-4.0 license.})}
\Address{Brandeis University, Waltham, United States of America}

\vfill

\begin{Abstract}
\noindent The production of jets and prompt isolated photons at hadron colliders provides stringent tests of perturbative QCD. The latest measurements performed by the ATLAS Collaboration at the LHC are presented in these proceedings. The inclusive prompt-photon production is measured for two distinct photon isolation cones, $R = 0.2$ and $0.4$, as well as for their ratio. This measurement is sensitive to gluon parton density distribution in the proton. In addition, a measurement of variables probing the properties of the multijet energy flow which are used to determine the strong coupling constant is presented. These measurements are compared to state-of-the-art NLO and NNLO predictions. Lastly, a measurement of new event-shape jet observables defined in terms of reference geometries with cylindrical and circular symmetries using the energy mover's distance is discussed.
\end{Abstract}

\vfill

\begin{Presented}
DIS2023: XXX International Workshop on Deep-Inelastic Scattering and
Related Subjects, \\
Michigan State University, USA, 27-31 March 2023 \\
     \includegraphics[width=9cm]{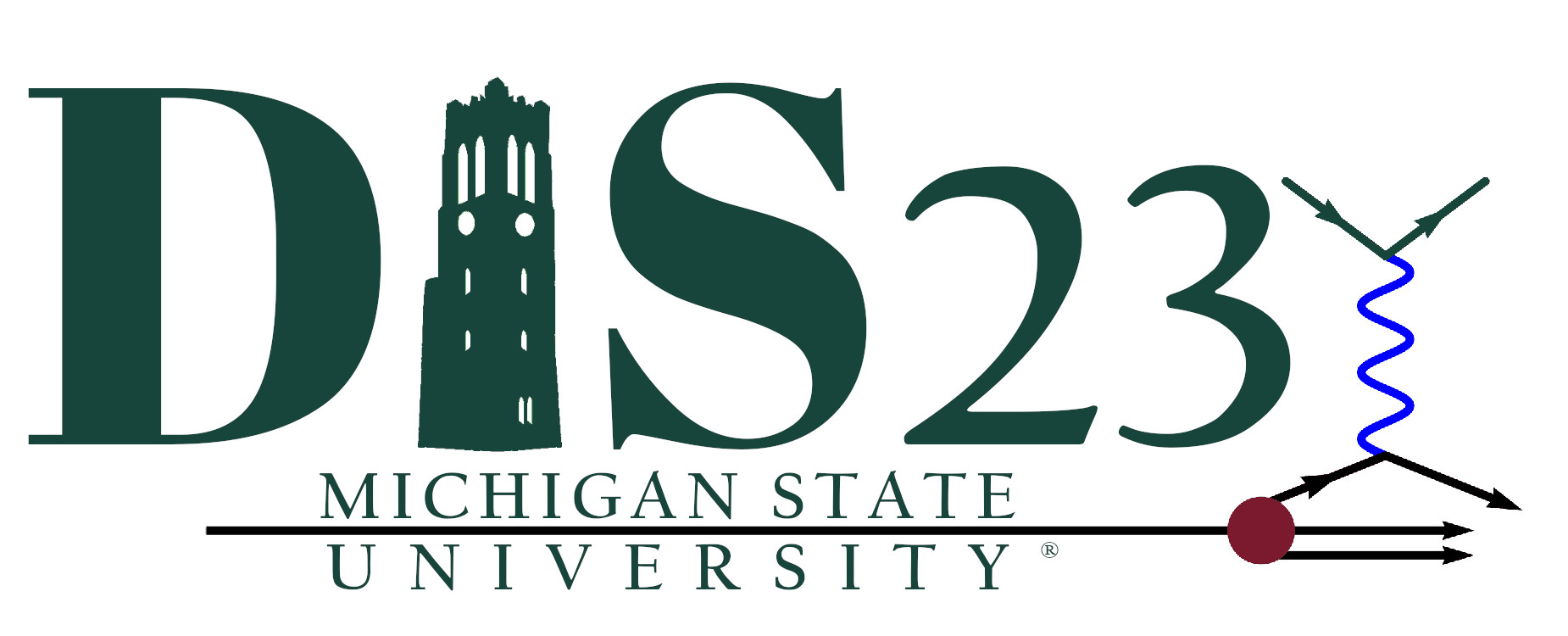}
\end{Presented}
\vfill

\end{titlepage}

\setlength{\parskip}{0.5em}

\section{Introduction}
\label{Sec:Intoduction}
The production of prompt photons\footnote{Photons that are not secondaries from hadron decays are considered as prompt.} with large transverse momenta ($\pt$) in proton-proton ($pp$) collisions, $pp \rightarrow \gamma + X$, provides a testing ground for perturbative QCD (pQCD). Since the dominant production mechanism in proton-proton collisions at the LHC proceeds via the $qg \rightarrow q\gamma$ process, measurements of prompt-photon production are sensitive to the gluon density in the proton.

In hadron colliders, due to the abundance of photons produced in neutral hadron decays and the contribution of fragmentation processes, prompt-photon production is studied by requiring photon isolation. This requirement is based on the amount of transverse energy ($\et$) allowed inside a cone of radius $R$ in the pseudorapidity-azimuth plane around the photon. For a detailed overview of the selection cuts employed in this measurement, please refer to Section 4 of Reference \cite{3}.

Multijet final states, produced in $pp$ collisions with large momentum transfer at the LHC, also provide an ideal testing ground for pQCD. Event shapes are a class of observables defined as functions of the final-state particles four momenta, which characterise the hadronic energy flow in a collision.

Transverse energy-energy correlations (TEEC) and their associated azimuthal asymmetries (ATEEC) are studied in Reference \cite{4}. The TEEC function is defined as the transverse-energy-weighted distribution of the azimuthal differences between jet pairs in the final state, while the ATEEC function is built as the difference between the forward ($\cos\phi > 0$) and backward ($\cos\phi < 0$) parts of the TEEC, as discussed in Section 1 of Reference \cite{4}. Both the TEEC and ATEEC functions are sensitive to gluon radiation and show a clear dependence on the strong coupling.

A novel class of event shapes, broadly called \textit{event isotropies}, was recently proposed to quantify the isotropy of collider events in terms of a Wasserstein distance metric. These distances are framed in terms of optimal transport problems, using the `Energy-Mover's Distance' (EMD), defined as \textit{the minimum amount of `work' necessary to transport one event $\mathcal{E}$ with $M$ particles into another $\mathcal{E}^{\prime}$ of equal energy with $M^{\prime}$ particles, by movements of energy $f_{ij}$ from particle $i \leq M$ in one event to particle $j \leq M^{\prime}$ in the other}, as shown in Equations (1) and (2) of Reference \cite{5}.

An event isotropy $\mathcal{I}$ is defined as a Wasserstein distance between a collider event $\mathcal{E}$ and a (quasi-)uniform radiation pattern $\mathcal{U}$, determined using the EMD: $\mathcal{I} \, (\mathcal{E}) = \mathrm{EMD} \, (\mathcal{E},\mathcal{U})$. The event isotropy $\mathcal{I}$ is bounded on $\mathcal{I} \in [0,1]$, where the least (most) isotropic events take values approaching $\mathcal{I} = 1$ $(\mathcal{I} = 0)$.

Three measurements performed at $\sqrt{s} = 13$~TeV with the $\thirdlumi$ of data recorded by the ATLAS experiment \cite{2} are presented in these conference proceedings.

\section{Inclusive-photon production}
\label{Sec:Anal1Inclusive}
Differential cross sections are measured as functions of the photon transverse energy, $\etgamma$, in different regions of the photon pseudorapidity, $\etagamma$, for $\etgamma > 250$~GeV and $\varetagamma < 2.37$. Photons are required to be isolated both at particle and detector levels using a `fixed cone' criterion: $E^{\mathrm{iso}}_{\mathrm{T}} < E^{\mathrm{iso}}_{\mathrm{T,cut}} = 4.2 \cdot 10^{-3} \cdot \etgamma +4.8$~GeV \cite{3}. The dependence of the cross section on the isolation cone radius, $R$, is investigated by measuring the ratios of the cross sections for $R = 0.2$ and $0.4$.

\subsection{Systematic uncertainties}

The dominant systematic uncertainties affecting the measurement of the differential cross sections arise from the uncertainty on the photon energy scale, the luminosity uncertainty, the uncertainty on the $R^{\mathrm{bg}}$ correlation, and the pile-up modelling. The total systematic uncertainty varies in the range $(3-20)\%$. The measurement of the ratios of the cross sections benefits from large cancellations for the uncertainties that are independent of the isolation cone radius. The total uncertainty is typically $<1\%$.

\subsection{Theoretical predictions}

The next-to-leading-order (NLO) pQCD calculations included in this measurement were computed using the programs \jetphox $1.3.1\_2$ and \sherpa $2.2.2$. The next-to-next-to-leading-order (NNLO) pQCD predictions are calculated in the \nnlojet framework. There are several differences between these calculations which are described in detail in Section 8 of Reference \cite{3}. The total uncertainty ranges from $\approx (10 - 15)\%$ for \jetphox, and it is $\approx 20\%$ for \sherpa. For the NNLO pQCD calculation of \nnlojet, the uncertainties are in the range $(1-6)\%$, being smaller than those in the NLO pQCD prediction by a factor $2-15$. The total uncertainty in the ratios of the cross sections benefits from large cancellations, being $\approx 1.5\%$ ($\approx 1\%$) for the predictions of \jetphox and \sherpa (\nnlojet).

\subsection{Results}

\begin{figure}
    \centering
    \begin{subfigure}[htpb]{0.475\textwidth}
        \centering
        \includegraphics[width=\textwidth]{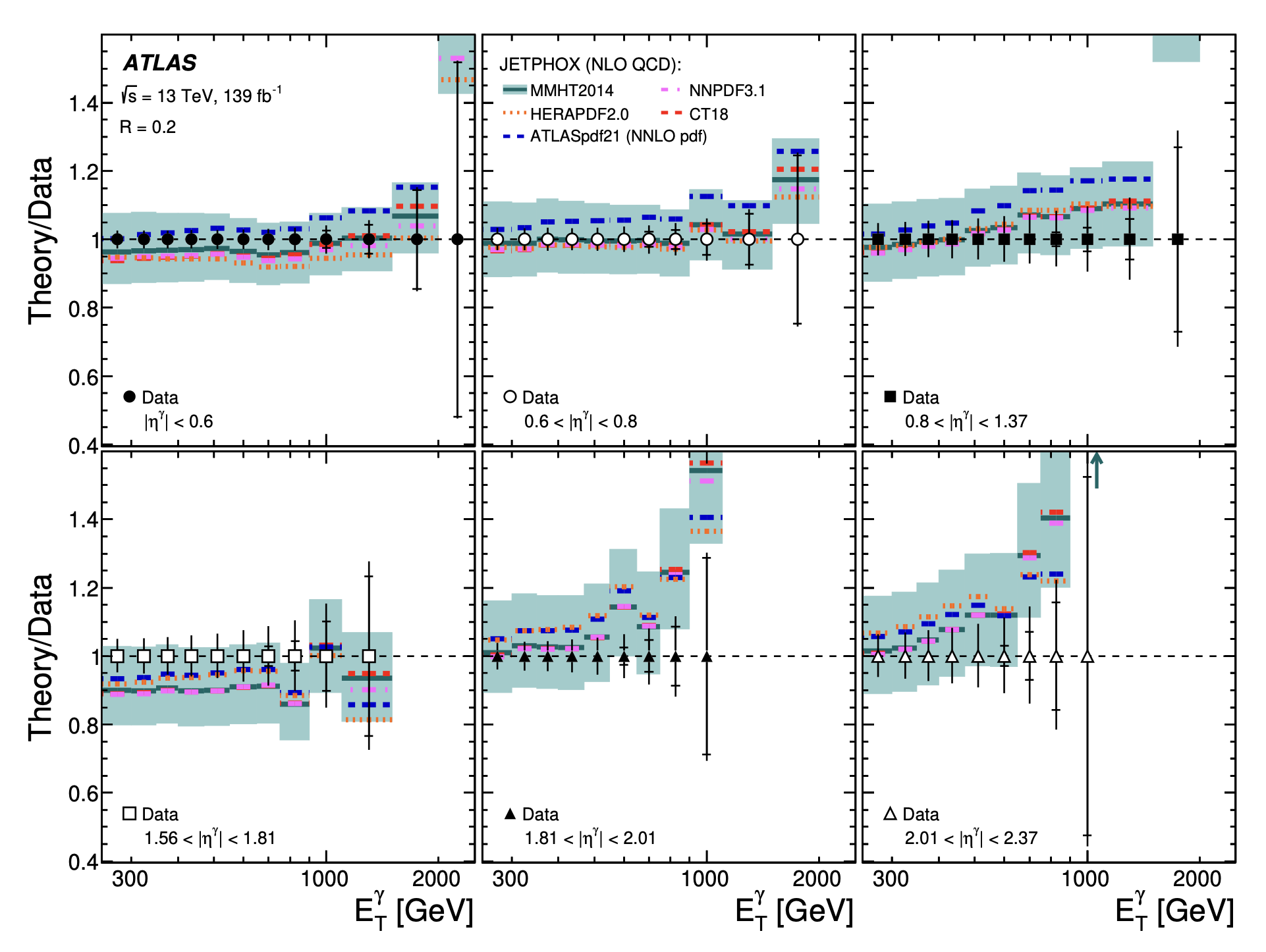}
        \caption{}
        \label{fig_1_1}
    \end{subfigure}
    \hfill
    \begin{subfigure}[htpb]{0.475\textwidth}
        \centering
        \includegraphics[width=\textwidth]{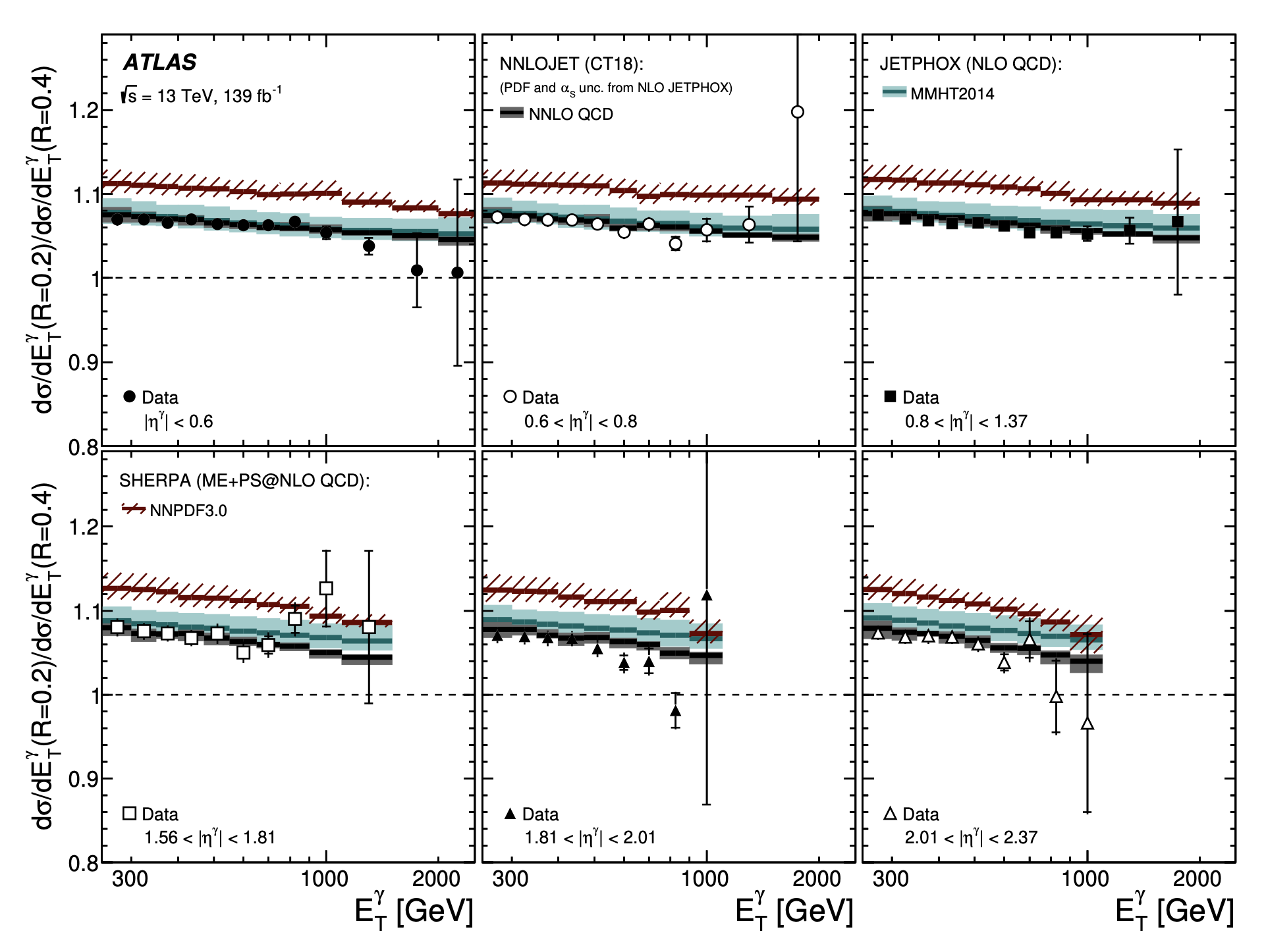}
        \caption{}
        \label{fig_1_2}
    \end{subfigure}
    \caption{Ratio of the NLO predictions from \jetphox based on different PDF sets and the measured differential cross sections with $R=0.2$ (a). Ratios of the measured differential cross sections for for $R=0.2$ and $0.4$ as functions of $\etgamma$ in different $\etagamma$ regions compared to the predictions from \sherpa, \jetphox, and \nnlojet (b) \cite{3}.}
    \label{fig_1}    
\end{figure}

The ratios of the predictions from \jetphox based on different PDFs and the data are shown in Figure \ref{fig_1_1} for $R=0.2$. The calculations of \jetphox are consistent with the measurements within uncertainties. Predictions based on MMHT2014, CT18 and NNPDF3.1 PDF sets are similar and the closest to the data for $\varetagamma < 1.37$ and $1.81 < \varetagamma < 2.37$. For $1.56 < \varetagamma < 1.81$, the predictions based on the HERAPDF2.0 PDF and ATLASpdf21 sets are the closest to the data.

The dependence of the inclusive photon cross section on $R$ is investigated by measuring the ratios of the cross sections for $R = 0.2$ and $0.4$, shown in Figure \ref{fig_1_2}. \sherpa predictions overestimate the data, while \jetphox gives a good description. The NNLO pQCD predictions give an excellent description of the data. These measurements provide a very stringent test of pQCD with reduced experimental and theoretical uncertainties, validating the underlying theoretical description up to $\mcal{O}(\alphas^2)$.

\section{Transverse Energy-Energy Correlations}
\label{Sec:Anal2TEEC}
Differential cross sections for the TEEC and ATEEC observables as functions of $\cos\phi$ are measured and compared with several Monte Carlo (MC) predictions inclusively and in ten bins of the scalar sum of the transverse momenta of the two leading jets, $H_{\mrm{T2}}$. The selected jets are required to have $\pt > 60$~GeV and $\vareta < 2.4$. The TEEC and ATEEC measured cross sections are compared to NNLO pQCD predictions for first time. A determination of the strong coupling constant is performed from this comparison.

\subsection{Systematic uncertainties}

The systematic uncertainties arising from the jet energy scale uncertainties dominate both the TEEC and ATEEC measurements, while the MC modelling and jet energy resolution uncertainties are also important for the TEEC and ATEEC, respectively. The total systematic uncertainty is of order $2\%$ ($1\%$) for the TEEC (ATEEC).

\subsection{Theoretical predictions}

The theoretical predictions for the TEEC and ATEEC functions are calculated at leading-order (LO), NLO and NNLO in pQCD. The NNLO pQCD corrections include real–real, real–virtual and virtual–virtual finite terms, as well as single- and double-unresolved terms and the finite remainder. The renormalisation and factorisation scales are set to the scalar sum of the transverse momenta of all final-state partons, $\muR = \muF = \hat{H}_{\mrm{T}}$. The uncertainties are computed from the scale variations, the PDF uncertainties, and those in the non-perturbative corrections to the pQCD predictions.

\subsection{Results}

The comparison of the measured cross sections for the TEEC and ATEEC functions with the predictions at different orders in pQCD are shown in Figures 6 and 7 of Reference \cite{4}. The description of the data provided by the NNLO pQCD predictions is excellent, improving with respect to the NLO prediction. The reduction of the scale uncertainties up to a factor of 3 is made evident from these figures, as well as the improvement in the description.

\begin{figure}[htpb]
    \centering
    \includegraphics[width=0.60\textwidth]{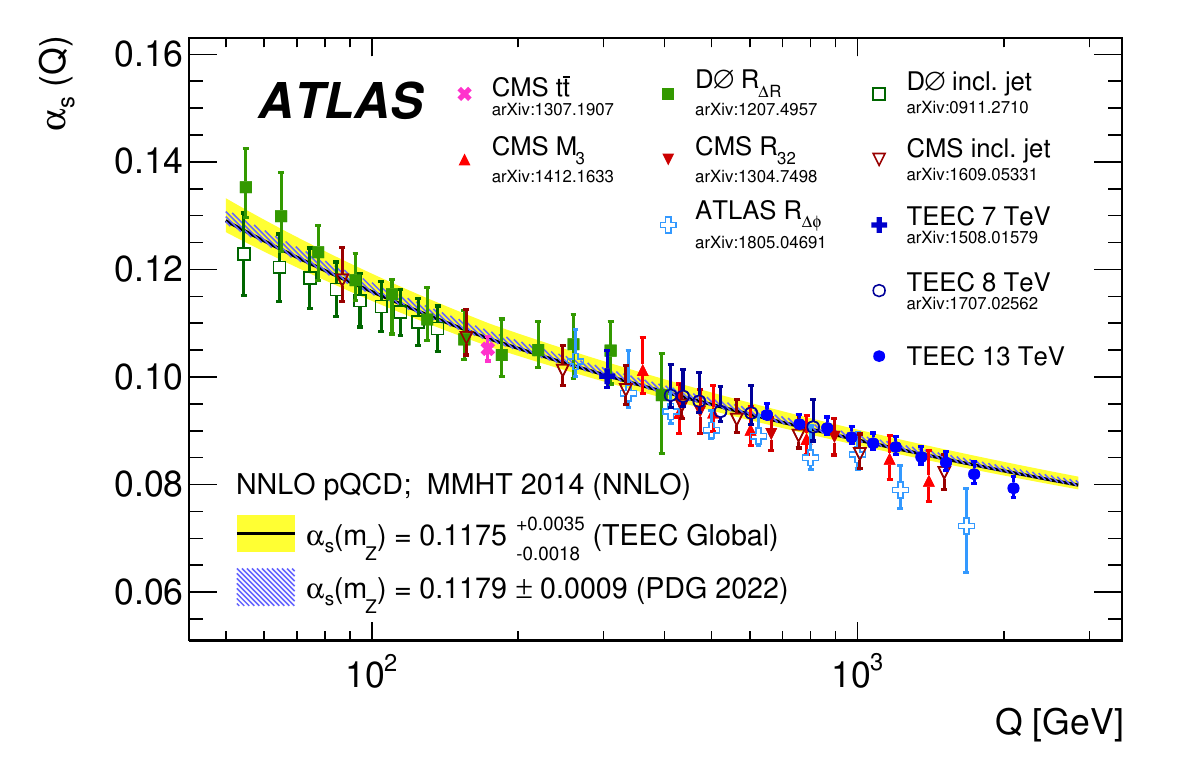}
    \caption{Comparison of the values of $\alphas(Q)$ determined from fits to the TEEC functions with the QCD prediction using the world average as input (hatched band) and the value obtained from the global fit (solid band) \cite{4}.}
    \label{fig_2_2}
\end{figure}

The strong coupling constant at the scale of the pole mass of the $Z$ boson, $\alphasmz$, was determined  from the comparison of the data to the theoretical predictions by means of a $\chi^2$ fit described in Section 9 of Reference \cite{4}, for both the TEEC and ATEEC distributions. To compare the results with other experiments, the value of the energy scale $Q$ is chosen as half of the average value of $\hat{H}_{\mrm{T}}$ for each $H_{\mrm{T2}}$ bin. Figure \ref{fig_2_2} shows the values of $\alphas(Q)$ from the fit to the TEEC distribution together with the world average band provided by the Particle Data Group \cite{6} and values of $\alphas$ obtained in other analyses. The results show a good agreement between all measurements and the renormalisation group equation prediction.

\section{Event isotropies using optimal transport}
\label{Sec:Anal3Isotropies}
The shape of the event isotropy observables $\mcal{I}^{N = 2}_\mrm{Ring}$, $\mcal{I}^{N = 128}_\mrm{Ring}$, and $\mcal{I}^{N = 16}_{\mrm{Cyl}}$ are measured in inclusive bins of $N_{\mrm{jet}}$ and $H_{\mrm{T2}}$. These observables are defined in Table 1 of Reference \cite{5}. The inclusive jet-multiplicity bins range from $N_{\mrm{jet}} \geq 2$ to $N_{\mrm{jet}} \geq 5$, and the inclusive bins of $H_{\mrm{T2}}$ are $H_{\mrm{T2}} \geq 500$~GeV, $H_{\mrm{T2}} \geq 1000$~GeV and $H_{\mrm{T2}} \geq 1500$~GeV. 

\subsection{Systematic uncertainties}

The dominant sources of systematic uncertainties are related either to the jet energy resolution or to the choice of MC model used in the unfolding. 

\subsection{Theoretical predictions}

Samples of MC simulated dijet and multijet events are used in this analysis. \pythia 8.230 is used as the nominal MC generator. Two sets of \sherpa 2.2.5 with the default AHADIC cluster hadronisation or with the \sherpa interface to the Lund string hadronisation model were used. Two sets of \herwig 7.1.3 multijet events were generated with the default cluster hadronisation model and either the default angle-ordered parton shower (PS) or alternative dipole PS. Two additional samples of dijet events with NLO matrix element accuracy were produced with \powhegbox, matched to either the \pythia 8 or angle-ordered \herwig 7 parton shower.

\subsection{Results}

The unfolded data are compared with predictions from several state-of-the-art Monte Carlo models. Good agreement is often observed between the LO and NLO Monte Carlo generators throughout the non-isotropic region of a given distribution; poorer agreement is seen as particle configurations become more isotropic.

\begin{figure}
    \centering
    \begin{subfigure}[htpb]{0.475\textwidth}
        \centering
        \includegraphics[width=\textwidth]{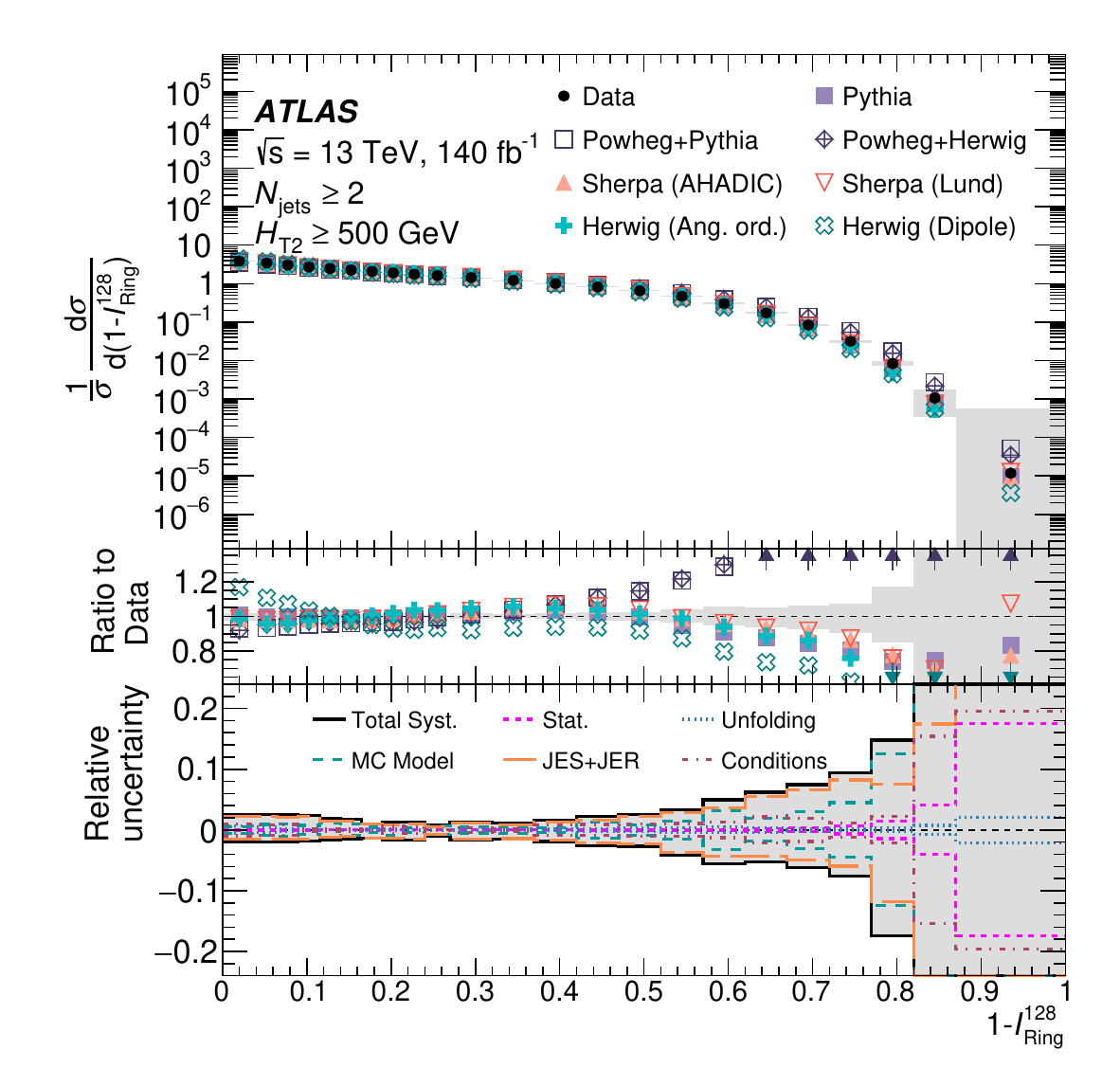}
        \caption{}
        \label{fig_3_1}
    \end{subfigure}
    \hfill
    \begin{subfigure}[htpb]{0.475\textwidth}
        \centering
        \includegraphics[width=\textwidth]{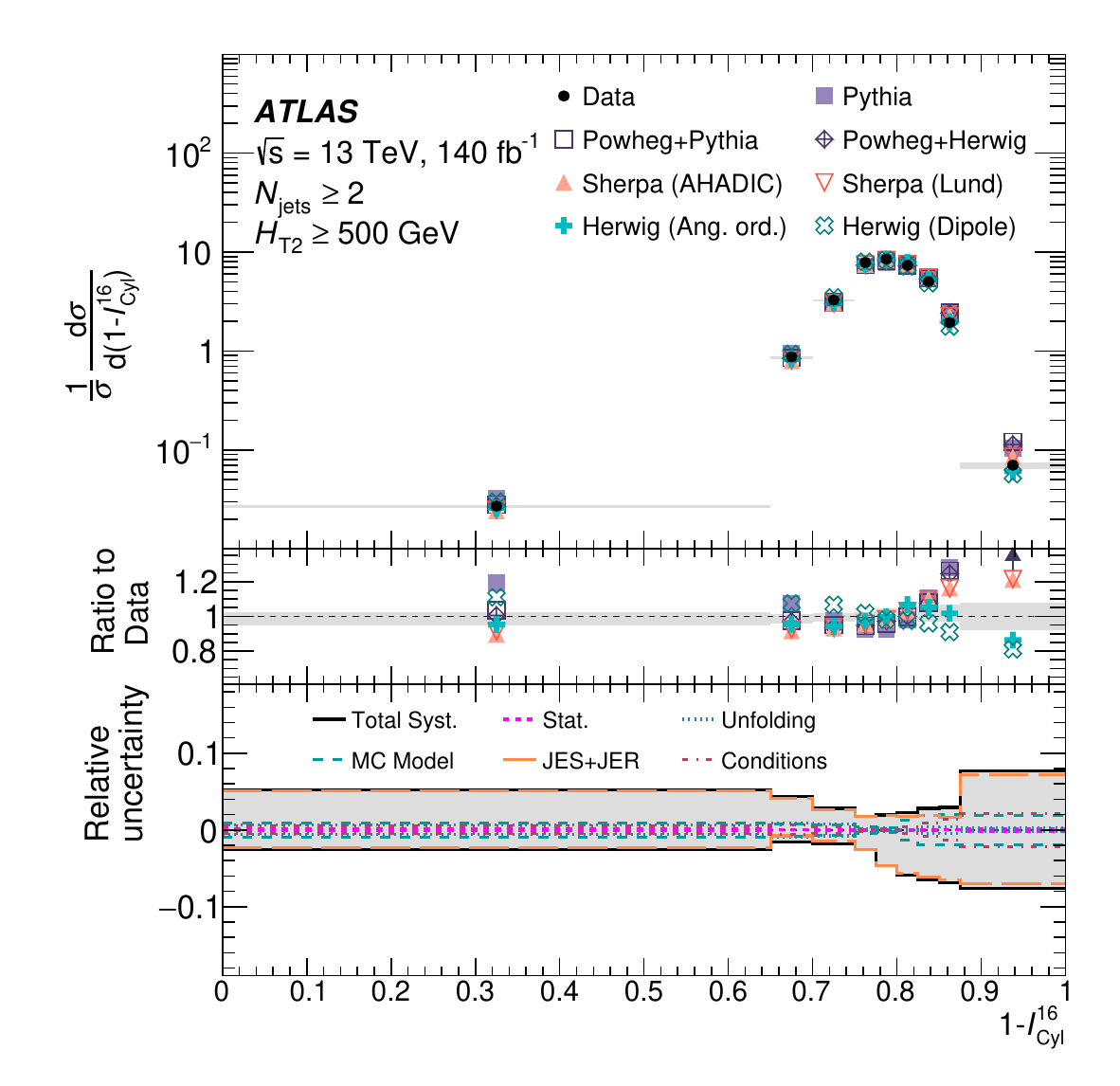}
        \caption{}
        \label{fig_3_2}
    \end{subfigure}
    \caption{The shape-normalized $\mcal{I}^{N = 128}_\mrm{Ring}$ and $\mcal{I}^{N = 16}_{\mrm{Cyl}}$ cross-sections are shown in (a) and (b), respectively. The data (closed circles) are compared with predictions from several MC generators. Events with $N_{\mrm{jet}} \geq 2$ and $H_{\mrm{T2}} \geq 500$~GeV are included \cite{5}.}
    \label{fig_3}    
\end{figure}

The most inclusive measurement of $1 - \mcal{I}^{N = 128}_\mrm{Ring}$ cross-sections is shown in Figure \ref{fig_3_1}. This distribution is saturated by well-balanced dijets events and by multijet events with isotropic configurations. The \powheg predictions are found to strongly disagree with those of the other MC generators. Large differences are also found between the \herwig angle-ordered and dipole shower models. No notable differences are seen between the \sherpa hadronisation models. The most inclusive measurement of $1 - \mcal{I}^{N = 16}_{\mrm{Cyl}}$ cross-sections is shown in Figure \ref{fig_3_2}. Multijet events that cover the rapidity–azimuth plane with activity in both the central and forward regions produce the highest values for this observable. None of the MC predictions accurately describe this observable. The predictions from the \pythia and \powheg samples are consistent except at low values, where \pythia overestimates the observed cross-section.

%
%

\addcontentsline{toc}{section}{Bibliography}

\end{document}